\documentclass[aps,pre,twocolumn,superscriptaddress]{revtex4-1}

\usepackage{amsmath}
\usepackage{amssymb}
\usepackage{graphicx}
  \usepackage{hyperref}
\usepackage[arrow, matrix, curve]{xy}
\usepackage{bm}
\usepackage{yhmath}
\usepackage{color}
\newcommand\diff{\mathrm{d}}

\usepackage[usenames,dvipsnames]{xcolor}
\hypersetup{colorlinks=true, linkcolor=BrickRed, urlcolor=blue!50!black, citecolor=blue!50!black}

\usepackage[normalem]{ulem}

\makeatletter
\newcommand\hide@visible[1]{%
  \bgroup\fboxsep=.3ex\colorbox{Gray}{begin hide}%
  #1\colorbox{Gray}{end hide}\egroup%
}
\newcommand\hide@hidden[1]{%
  \bgroup\fboxsep=.3ex\colorbox{Gray}{hidden text}%
}
\newcommand\hide@invisible[1]{}
\newcommand\makevisible{\let\hide\hide@visible}
\newcommand\makehidden{\let\hide\hide@hidden}
\newcommand\makeinvisible{\let\hide\hide@invisible}
\makeatother
\makehidden

\begin{document}


\title{Strongly confined fluids: Diverging time scales and slowing down of  equilibration}


\author{Rolf Schilling}
\affiliation{Institut f\"ur Physik, Johannes Gutenberg-Universit\"at Mainz,
 Staudinger Weg 9, 55099 Mainz, Germany}


\date{\today}

\begin{abstract}
The Newtonian dynamics of strongly confined  fluids exhibits a rich behavior. Its confined and unconfined degrees of freedom  decouple for confinement length $L \to 0$.
In that case and for a slit geometry the intermediate scattering functions $S_{\mu\nu}(q,t)$ simplify,  resulting  for $(\mu,\nu) \neq (0,0)$ in a Knudsen-gas like behavior of the confined degrees of freedom, and otherwise in $S_{\parallel}(q,t)$,  describing the structural relaxation of the unconfined ones. Taking the coupling into account we 
prove that the energy fluctuations relax exponentially. For smooth potentials the relaxation times diverge as  $L^{-3}$ and  $L^{-4}$, respectively, for the confined and unconfined degrees of freedom. The strength of the $L^{-3}$ divergence can be calculated analytically. It depends on the pair potential and the two-dimensional pair distribution function. Experimental setups are suggested to test these predictions.
\end{abstract}






\maketitle


\section{Introduction}\label{Sec: intro}
Compared to calculating static properties of macroscopic systems in thermodynamic equilibrium
the calculation of dynamical quantities like, e.g., the time evolution of correlation functions
in equilibrium or of quantities in nonequilibrium is much harder. A prominent example is the intermediate scattering function $S(q,t)$ of a simple classical fluid \cite{HMcD06}. Whereas its short-time
behavior can be deduced from the computation of static correlation functions, a reliable calculation
of its decay on intermediate and long-time scales is not possible, in general,  due to the interactions between the particles.

There are very few macroscopic classical systems with Newtonian dynamics for which exact results exist. Without attempting  completeness we mention some examples. For the ideal bulk gas of point particles the self intermediate scattering function $S^{(s)}(q,t)$
is known analytically \cite{HMcD06}. The linear correction in density has also been determined  \cite{MdeSch01}. The evolution of the nonequilibrium
probability distribution function in the $N$-particle phase space of the Knudsen gas (an ideal gas
confined between flat, parallel and hard walls) was determined \cite{HL68,L74} and of correlation functions for a harmonic crystal \cite{MM60}. These examples refer to systems
without particle-particle interactions or to systems which can be transformed to noninteracting modes,
as in Ref. \cite{MM60}. Analytically exact results for the dynamics of interacting particles are particularly rare.
An example is a fluid of hard rods in one dimension for which, e.g., the long-time decay of the velocity autocorrelation function was determined \cite{J65}.

But, there exist limiting cases which allow predictions of the dynamical properties.
A famous example is the hydrodynamic limit for a fluid leading to the Navier-Stokes equations. For a general discussion see the recent review \cite{H16}.
Another well-known example is Brownian motion. Starting from the {\it microscopic} dynamics of a
particle with mass $M$ immersed in a solvent of particles having mass $m$ the dynamics of the Brownian
particle is described by a Langevin equation in case of $m/M \ll 1$ \cite{F75}. There are more
situations for which a smallness parameter exists, like $m/M$ for a Brownian particle, and for
which a kinetic equation can be derived from the microscopic dynamics (see the review \cite{Sp80} and references therein).
Such a typical situation occurs  for two subsystems interacting weakly with a coupling constant $\lambda$. If $H_0=H_1+H_2$ is the (classical) Hamiltonian
of the unperturbed system and $\lambda H_{\rm int}$ its
perturbation, the time evolution of a phase space function $A(x)$ is given by $A(x(t))=\exp (i \mathcal{L} t)
A(x(0))$. Here $x(t)$ is the trajectory in phase space and $\mathcal{L}=\mathcal{L}_0 + \lambda \mathcal{L}_{\rm int}$,
the Liouville operator. $\mathcal{L}_0$ and $\mathcal{L}_{\rm int}$ correspond to $H_0$ and $H_{\rm int}$, respectively.
The Taylor expansion of $\exp(i \mathcal{L} t)$ up to, e.g., second order in $t$ allows only to
calculate the short-time dynamics of the correlation function $\langle A(x(t)) A(x(0))\rangle$ where $\langle (\cdot )\rangle$
denotes canonical averaging over the initial conditions $x(0)$ in phase space.
The long time behavior can only be obtained by summing up an infinite number of higher order terms. This was done  first for
a quantum system for which a relaxation time $\tau$ for the 
approach-to-equilibrium  was found diverging
as $\lambda^{-2}$ \cite{vH55}. Taking the so-called van-Hove-limit $t \rightarrow \infty$ and $\lambda \rightarrow 0$ such that
$\tilde{t}= \lambda^2 t ={\rm const}$ one obtains a kinetic equation describing the approach-to-equilibrium on a time
scale $\tilde{t}$. Correction terms to that limit were also determined \cite{L86}. Similar perturbational
treatments for classical systems were elaborated in Ref. \cite{PB59}.

Confined fluids interpolate between bulk and 2D (or 1D) fluids. Their dynamics
is influenced by the confinement and its geometry.  Therefore the question raises how the fluid dynamics will change  if the confinement becomes stronger and stronger. In that case the fluid becomes  quasi-two- (or quasi-one-) dimensional.
Recently, it was shown that thermodynamic and structural properties of fluids strongly confined by two flat,
parallel and hard walls can be calculated perturbatively, with $n_0L^2$ as a smallness parameter \cite{FLS12,LFS14}.
$L$ is the accessible width in transversal direction and $n_0=N/A$ is the 2D number density for $N$ identical
particles and a wall area $A$. This perturbational approach to calculate static thermodynamic quantities is based on the
observation that the particle's lateral and transversal degrees of freedom (d.o.f.) become decoupled for
$L \rightarrow 0$. Following Ref. \cite{vH55} one would expect that the equilibration time $\tau$ of  an initial state  in which the transversal and lateral d.o.f. are out-of-equilibrium
should diverge as $(n_0L^2)^{-2}  \sim L^{-4}$ for $L \rightarrow 0$. Similar behavior would be expected to occur for fluids in narrow cylindrical
tubes with diameter $R$ and in narrow two-dimensional channels with width $W$  for $R \rightarrow 0$ and $W \rightarrow 0$, respectively.

It is the main object of the present work to elaborate on the relaxational behavior of fluids in \textit{strong} confinement. Not much is known for this kind of problem. There have been investigations  of  moderately confined fluids.  On the experimental side  the influence of the confinement on  the molecular relaxation was explored\cite{KD89}.   Using confocal  microscopy for hard sphere suspensions the mean square displacements and the diffusion constants were determined as a function of the wall separation \cite{NEPW07,EEMD09,EMD11}. The same was done by computer simulations \cite{AIMPS84,MTEH08,MLGORFV14}. Many of these activities were motivated by studying transport properties of dense liquids and  the structural glass transition.
    
In the present article we will show  that the approach-to-equilibrium of fluids in strong confinement slows down with decreasing confinement length. The slowing-down is much stronger for the lateral than for the transversal d.o.f.. 
This follows for \textit{smooth}  pair potentials from the existence of  two \textit{different}  time scales related to the energy relaxation of the lateral and transversal (d.o.f.) diverging for vanishing confinement length .  The outline of our paper is as follows. The next section, Sec. II, presents the model, and the leading order dynamics in $L$ is discussed in Sec. III. The main result, i.e., the derivation of an exact equation of motion for the energy correlator of the transversal and lateral d.o.f. will be presented in Sec. IV. In particular, this section derives the power law divergencies of the correponding relaxation times. Finally, Sec. V contains a summary and some conclusions. In order not to impede the reading of the main text  technical details are presented in Appendices.

\section{Model}\label{Sec: model}

We consider $N$ identical point particles with mass $m$ confined in a three-dimensional domain $\mathcal{D}$.
Its classical Hamiltonian reads

\begin{equation} \label{eq1}
H \Big(\{\vec{p}_i \}, \{\vec{x}_i \} \Big) =\sum\limits_{i=1}^N \Big[\frac{1}{2m} \vec{p}_i^2 + U\Big(\vec{x}_i \Big)\Big] +
\frac{1}{2} \sum\limits_{i \neq j} \upsilon \Big(|\vec{x}_i - \vec{x}_j| \Big)
\end{equation}
where $\vec{p}_i$ and $\vec{x}_i$ is the momentum and position, respectively, of the ith particle.
$\upsilon (x_{ij}=|\vec{x}_i - \vec{x}_j|)$ is a \textit{smooth}  central symmetric pair potential and

\begin{align}
\label{eq2}
U(\vec{x}) =
\begin{cases}
\infty \quad , \quad \vec{x} \notin \mathcal{D}\\
0 \quad, \quad \vec{x} \in \mathcal{D}
\end{cases}
\end{align}
is a hard-wall potential confining the particles to the domain $\mathcal{D}$.
\textit{Smooth} means that the pair potential's derivatives exist up to arbitrary order for all $x_{ij}$, except for $x_{ij}=0$. This condition on the potential $v(x_{ij})$ is fulfilled for, e.g., a Coulomb and a Lennard-Jones potential. It mainly excludes hard-core interactions with a finite hard-core size.
Note, that the walls are neutral and reflect particles elastically.
The extension of our results to walls with additional  particle-wall interactions is straightforward.
We call the confinement  of a three-dimensional fluid  \textit{strong} if  the particles form only a single monolayer and a single  chain-like particle arrangement for a slit and a tube or a 2D channel geometry, respectively. This situation can occur because a realistic pair potential, e.g., a Lennard-Jones potential, becomes more and more repulsive with decreasing distance $x_{ij}$. For charged particles it is even purely repulsive for all $x_{ij}$. Accordingly, the statistical weight of configurations with
lateral distance $r_{ij}=0$  for a given pair $(i,j)$ of particles is of the order $\exp[-(l/L)^m]$. The microscopic length $l$ and the positive parameter $m$ characterize the divergence of $v(x_{ij})/k_BT$ at  $x_{ij}=0$. For a Coulomb and a Lennard-Jones potential it is $m=1$ and $m=12$, respectively. Since our perturbational approach leads to an expansion in powers of $L^2$ these exponentially small contributions can be neglected for $L$ small enough.
This condition allows to expand the pair potential into a Taylor
series of the confined coordinates called transversal d.o.f. in the following. The unconfined coordinates comprise the lateral d.o.f.. Let us restrict to a
slit geometry with a width $L$ and a wall area $A$. The perturbational procedure for a tube or a 2D channel can be done analogously. Decomposing
$\vec{x}_i=(\vec{r}_i, z_i)$ into lateral and transversal coordinates $\vec{r}_i = (x_i, y_i)$ and
$z_i$, respectively, and following Refs. \cite{FLS12,LFS14} one obtains for the pair interaction energy

\begin{eqnarray} \label{eq3}
V\Big(\{\vec{x}_i \}\Big)&=&\frac{1}{2} \sum\limits_{i \neq j} \upsilon \Big(\vec{x}_i-\vec{x}_j\Big) \nonumber \\
&& =V \Big(\{\vec{r}_i \}\Big) + V_{\parallel, \bot} \Big(\{\vec{r}_i \}, \{z_i \} \Big)
\end{eqnarray}
where

 \begin{equation} \label{eq4}
 V_{\parallel, \bot} \Big(\{\vec{r}_i \}, \{z_i \} \Big) = \sum\limits_{\nu=1}^{\infty} \frac{1}{2}\sum\limits_{i \neq j} \upsilon_ \nu \Big(r_{ij} \Big)
 \Big( z_i - z_j \Big)^{2\nu}
 \end{equation}
is the interaction energy of the lateral  with the transversal  d.o.f..
 The $\nu$th order coefficients are related to the derivatives $\upsilon'(r_{ij})$, $\upsilon^{''}(r_{ij})$, etc. of
the pair potential

\begin{eqnarray} \label{eq5}
&& \upsilon_1(r_{ij})=\upsilon'(r_{ij})/2 r_{ij} \nonumber\\
&&\upsilon_2(r_{ij})=[\upsilon{''}(r_{ij}) r_{ij} - \upsilon{'}({r_{ij}})] / 8 r^3_{ij}
\end{eqnarray}
etc.. Since $z_i=\mathcal{O}(L)$, Eq.~(\ref{eq4}) corresponds to an expansion with respect to
$L$. Note that this expansion requires the smoothness of the pair potential. Making use of Eqs.~(\ref{eq1}) - (\ref{eq4}) and of $\vec{p}_i = (\vec{P}_i, \vec{P}_i^z)$
we get the corresponding Liouvillean

\begin{equation} \label{eq6}
\mathcal{L}=\mathcal{L}_0+\mathcal{L}_1  \quad , \ \mathcal{L}_0=\mathcal{L}_0^{\parallel} + \mathcal{L}_0^{\bot}
 \end{equation}
with the unperturbed parts

\begin{eqnarray} \label{eq7+8}
\mathcal{L}^{\parallel}_0&=&-i \sum\limits_{j=1}^N \Big[ \frac{1}{m} \vec{P}_j \cdot \frac{\partial }{\partial \vec{r}_j} -
\frac{\partial V}{\partial \vec{r}_j} \cdot \frac{\partial}{\partial \vec{P}_j} \Big] \nonumber \\
\mathcal{L}^\bot_0&=&-i \sum\limits^N_{j=1} \frac{1}{m} P^z_j \frac{\partial}{\partial z_j} =\sum\limits_{j=1} ^N \mathcal{L}^\bot_{0,j}
\end{eqnarray}
and the perturbation $\mathcal{L}_1=\mathcal{L}^{\parallel}_1 + \mathcal{L} ^\bot_1$ where

\begin{eqnarray} \label{eq9+10}
\mathcal{L}^{\parallel}_1&=& i\sum^\infty_{\nu=1} \sum\limits_{i \neq j} \Big( z_i-z_j\Big)^{2 \nu}
\frac{\upsilon'_\nu(r_{ij})}{r_{ij}} \vec{r}_{ij} \cdot \frac{\partial}{\partial \vec{P}_i} \nonumber \\
\mathcal{L}^\bot_1&=& i \sum\limits_{\nu=1}^\infty 2 \nu \sum\limits_{i \neq j} \Big(z_i -z_j\Big) ^{2 \nu -1} \upsilon_{\nu} \Big(r_{ij}\Big)
\frac{\partial}{\partial P^z_i}  \ .
\end{eqnarray}

The $\nu$th order term of $\mathcal{L}^{\parallel}_1 $ and $\mathcal{L}^\bot_1$ is of order $L^{2 \nu}$ and
$L^{2\nu-1}$, respectively. Accordingly, the dynamics of the strongly confined fluid in leading order is reduced to
the dynamics of the decoupled lateral and transversal d.o.f. which will be discussed in the following section. Taking into
account their coupling will lead to a "kinetic" equation derived in Sec. IV

\section{Leading order dynamics}\label{Sec: leadingdyn}

The intermediate scattering function of a fluid is of theoretical and of experimental interest. For a fluid in slit
geometry it is an infinite-dimensional matrix with matrix elements

\begin{equation} \label{eq11}
S_{\mu \nu} (q,t) =\frac{1}{N} \sum\limits_{m,n=1}^N \langle e^{-i\vec{q}\cdot [\vec{r}_m (t)-\vec{r}_n(0)]} e^{- i[Q_\mu z_m (t)-Q_\nu z_n (0)]}\rangle
\end{equation}
with the 2D wave vector $\vec{q}=(q_x, q_y)$, $Q_\mu=2 \pi \mu/L$ and $\mu$ an integer.  For more details see
Refs. \cite{LBOHFS10,LSKF12}. The angular brackets denote  canonical averaging over the initial conditions
$(\{\vec{r}_n(0), z_n(0)\}$, $\{\vec{P}_n(0)$, $P^z_n (0) \})$. 
 In leading order  in $L$ this canonical average factorizes into the product $\langle \rangle ^{\parallel} \langle \rangle ^\bot$, involving the
canonical averages over the lateral $(\{\vec{r}_n (0)\}$, $(\{\vec{P}_n (0)\})$ and over the
transversal $(\{z_n (0)\}$, $\{P^z_n (0)\})$ d.o.f. \cite{FLS12}. Furthermore,  in leading order  in $L$ the time evolution operator $\exp{(i\mathcal{L}t)}$ factorizes   
into $\exp{(i\mathcal{L}_0^{\parallel}t)}$ $\exp{(i\mathcal{L}_0^{\perp}t)}$.
Then Eq.~(\ref{eq11}) implies
that $\langle \exp (-i\vec{q}\cdot[\vec{r}_m (t)-\vec{r}_n(0)]) \rangle ^{\parallel}$  and $\langle \exp (-i[Q_\mu z_m (t)-Q_\nu z_n (0)]) \rangle ^\bot$ have to be calculated, where the time evolution is generated by $\mathcal{L}_0^{\parallel}$ and $\mathcal{L}_0^{\perp}$, respectively.  Therefore, for $(\mu,\nu) = (0,0)$ one obtains $S_{00} (q,t) \simeq S_{\parallel}(q,t)$,  the intermediate scattering function of the 2d fluid of the unperturbed lateral d.o.f.. 
In leading order in $L$ there is no interaction between the transversal d.o.f.. Consequently, they form an ideal gas where
$z_i(t)$ is confined between $-L/2$ and $L/2$, which represents a
one-dimensional Knudsen gas. Then the transversal correlators above for $(\mu, \nu) \neq (0,0) $  are nonzero
for $m=n$, only. Taking this into account we get from Eq.~(\ref{eq11})  in leading order for $(\mu, \nu) \neq (0,0) $

\begin{equation} \label{eq12}
S_{\mu \nu} (q,t) \simeq S^{(s)}_{\parallel} (q,t) S^{(K)}_{\mu \nu} (t)
\end{equation}
with the self part of the 2D  intermediate scattering function
$S_{\parallel}^{(s)} (q,t)= \frac{1}{N}\sum\limits_{n=1}^N \langle e^{-i \vec{q} \cdot [\vec{r}_n(t)-\vec{r}_n(0)]} \rangle^{\parallel}$ and the
"Knudsen" correlators

\begin{equation} \label{eq13}
S_{\mu \nu}^{(K)} (t) =\langle \exp (-i[Q_\mu z_n(t)-Q_\nu z_n (0)])\rangle^\bot \quad ,
\end{equation}
which do not depend on the particle index $n$.

Using the result of Ref. \cite{L74} $S^{(K)}_{\mu \nu} (t)$ can be calculated exactly. One obtains for $(\mu, \nu) \neq (0,0)$ (see Appendix A)

\begin{eqnarray} \label{eq14}
&& S^{(K)}_{\mu \nu} (t)=(-1)^{\mu + \nu} \Big\{\exp [-8 \pi^2 \mu^2(t/t_K(L))^2](\delta_{\mu,- \nu}+\delta_{\mu \nu}) + \nonumber\\
&& + \sum\limits_{k=0}^\infty c^k_{\mu \nu} \exp [-2 \pi^2 (2k+1)^2 (t/t_K (L))^2 ]\Big\}
\end{eqnarray}
with the "Knudsen" time scale

\begin{equation} \label{eq15}
t_K(L)=2 L/\upsilon_{th} \quad , \ \upsilon_{th} =(k_BT/m)^{1/2}
\end{equation}
the time for bouncing back and forth of a particle with thermal velocity $\upsilon_{th}$.
The coefficients $c^k_{\mu\nu}$ are given in Appendix A. The long time decay is $S^{(K)}_{\mu \nu}(t)
\simeq  c^0_{\mu \nu} \exp [-2 \pi^2 (t/t_K (L))^2 ]$, i.e. a Gaussian
decay on a time scale $t_K(L) \sim L$. For argon ($^{40}_{80}) Ar$ at room temperature and $L=1nm$ one
obtains $t_K(1nm) \cong 8.10 \times 10 ^{-12}$ sec. $t_K(L)$ is much smaller than the corresponding
relaxation time $\tau_{\parallel}$ of $S_{\parallel}^{(s)} (q,t)$. Therefore it follows with Eq.~(\ref{eq12}) and $S_{\parallel}^{(s)} (q,0)=1$

\begin{align}
\label{eq16}
S_{\mu \nu} (q,t) \simeq
\begin{cases}
S_{\parallel} (q,t) \ , & (\mu, \nu)=(0,0) \\
S^{(K)}_{\mu \nu} (t) \ , & (\mu, \nu) \neq (0,0)  \ ,
\end{cases}  
\end{align}
i.e., the intermediate scattering functions in leading order in $L$  separate into
the intermediate scattering function of the unperturbed 2D fluid and the corresponding correlators of a Knudsen gas.

\section{Energy relaxation}\label{Sec: kineticeq}

In a final step we investigate the influence of the coupling between the
lateral and transversal d.o.f. on the relaxational behavior. For this we
observe that the total energy of the lateral and of each of the transversal d.o.f.
are conserved in leading order in $L$ \cite{FLS12}. Accordingly, these quantities are slow variables for finite but small $L$.
Let $h_\bot (P^z_i) = (P^z_i)  ^2/ 2m$ be the kinetic energy of the ith particle's transversal
d.o.f. and $\delta h_\bot (P^z_i)=h_\bot (P^z_i) - \langle h_\bot (P^z_i)\rangle$ its
fluctuation. Similarly, $\delta H_{\parallel} (\{\vec{p}_i\}, \{\vec{r}_i \}) =H_{\parallel} (\{\vec{p}_i\}, \{\vec{r}_i \}) -\langle H_{\parallel} \rangle$ denotes the fluctuation of the  total energy $H_{\parallel}(\{ \vec{P}_i\},
\{ \vec{r}_i \})=\sum\limits_i \vec{P}^2_i / 2m + V (\{\vec{r}_i \})$ of the lateral
d.o.f..  In the following we derive an equation of motion for the tranversal self correlator
$C^{(s)}_\bot (t)= \frac{1}{N} \sum\limits_i \langle \delta h_\bot (P^z_i (t)) \delta
h_\bot (P^z_i(0))\rangle$ and for the lateral correlator $C_{\parallel}(t)= \langle \delta H_{\parallel}
(\{\vec{P}_i(t)\}$, $\{\vec{r}_i (t) \})$ $\delta H_{\parallel} (\{\vec{P}_i(0)\}, \{\vec{r}_i(0)\}) \rangle$.
In principle, one could perform a perturbational calculation by expanding the evolution operator
$\exp[i (\mathcal{L}_0 + \mathcal{L}_1)t]$ with respect to $\mathcal{L}_1$, similar to the
procedure in Refs. \cite{vH55,PB59}. This would require to sum up an infinite number of terms
(diagrams). Here we demonstrate that the Zwanzig-Mori projection formalism \cite{HMcD06,F75} is the most suitable  method to derive an exact equation of motion for $C^{(s)}_\bot (t)$ and $C_{\parallel} (t)$ in the limit $L \to 0$. 

Let us just focus on
$C^{(s)}_\bot (t)$. With the projector $\mathcal{P}_\bot = |\delta h_\bot (P^z_s) \rangle \langle \delta h_\bot (P^z_s)|
/\langle (\delta h_\bot (P^z_s))^2 \rangle$ for a ``tagged'' particle $s$ one obtains the Mori equation
\cite{HMcD06,F75}

\begin{equation} \label{eq17}
\dot{C}^{(s)}_\bot (t) + \int\limits_0^t d t' K_\bot^{(s)} (t-t') C_\bot^{(s)} (t')=0
\end{equation}
with the memory kernel

\begin{eqnarray} \label{eq18}
K^{(s)}_\bot (t) &=& \langle \delta h_\bot (P^z_s) \mathcal{L}\mathcal{Q}_\bot \exp [-i \mathcal{Q}_\bot \mathcal{L} \mathcal{Q}_\bot t] \mathcal{Q}_\bot
\mathcal{L} \delta h(P^z_s)\rangle \times \nonumber \\
&& \times \langle  (\delta h_\bot (P^z_s))^2 \rangle^{-1}
\end{eqnarray}
and the projector $\mathcal{Q}_\bot = 1-\mathcal{P}_\bot$. The r.h.s. of Eq.~(\ref{eq18}) simplifies because $\mathcal{Q}_\bot \mathcal{L} \delta h_\bot (P^z_s)\equiv \mathcal{L} h_\bot (P^z_s)=
\mathcal{L}^\bot_1 \delta h_\bot (P^z_s)$ since $\mathcal{L} \delta h_\bot (P^z_s)$ is orthogonal to $\delta h_\bot (P^z_s)$.
Furthermore $\mathcal{L}_0 \delta h_\bot (P^z_s)=0$ and $\mathcal{L}^{\parallel}_1 \delta h_\bot (P^z_s)=0$ has been used. The former holds since $\delta h_\bot (P^z_s)$ is conserved under the time
evolution generated by the unperturbed Liouvillean $\mathcal{L}_0$ and the latter follows immediately with use of Eq. (\ref{eq9+10}). Because $\mathcal{L}^\bot_1 \delta h_\bot (P^z_s)$ is of
order $L$ (cf. Eq.~(\ref{eq9+10})) one can replace $Q_\bot \mathcal{L} Q_\bot$ in the exponent of Eq.~(\ref{eq18}) by the zero order term $\mathcal{Q}_{\perp}^{(0)}\mathcal{L}_0 \mathcal{Q}_\bot^{(0)}$ and 
$\langle(\cdot)\rangle $ by $\langle(\cdot) \rangle^{(0)}$, where the latter denotes the canonical average with respect to the unperturbed Hamiltonian $H_0$  (see also Ref.\cite{FS86}).
Here it is  $\mathcal{Q}^{(0)}_\bot=1-\mathcal{P}^{(0)}_\bot$ and
$\mathcal{P}^{(0)}_\bot =|\delta h_\bot (P^z_s) \rangle^{(0)(0)} \langle \delta h_\bot (P^z_s)|/ \langle(\delta h_\bot (P^z_s)^2 \rangle^{(0)}$. Then the leading order result for the kernel becomes (see Appendix B)

\begin{equation} \label{eq19}
K^{(s)}_\bot \simeq  - k_1 C^\bot_1 (t) \ddot{C}_1 (t) - k_2 \ddot{C}^\bot_2 (t)
\end{equation}
where $k_1$ and $k_2$ involve static correlators of the unperturbed lateral system, only. $\dot{f}(t)$ denotes the derivative of $f(t)$ with respect to $t$. The transversal correlators appearing in Eq.~(\ref{eq19}) are given by $C^\bot_m(t) = \langle (z_s(t))^m (z_s (0))^m \rangle ^\bot$,
$m=1,2.$ It is important to note that the result (\ref{eq19}) for $K^{(s)}_\bot (t)$ is not valid for a \textit{hard-core} pair
potential because $k_1$ and $k_2$ involve static correlation functions of $\upsilon_1(r)$ which is proportional to the
derivative $\upsilon'(r)$ of the pair potential (cf. Eq.~(\ref{eq5})). Using scaled variables $\tilde{z_s}=z_s/L$ and $\tilde{t} =t/t_{K}(L)$ with
$t_{K}(L)$ from Eq.~(\ref{eq15}) one obtains

\begin{equation} \label{eq20}
C^\bot_m (t) =L^{2m} \tilde{C}_m^\bot (\tilde{t}) \quad .
\end{equation}

The correlators $\tilde{C}^\bot_m(\tilde{t})$ are independent on $L$ and can be calculated analytically (see Appendix A). Making use of
the scaling relation, Eq.~(\ref{eq20}), we obtain from Eq.~(\ref{eq19})

\begin{equation} \label{eq21}
K^{(s)}_\bot (t) \cong \Big[L^4/t_K (L)^2 \Big] \tilde{K}_\bot^{(s)} (\tilde{t})
\end{equation}
where the $L$-independent relaxation kernel $\tilde{K}_\bot^{(s)} (\tilde{t})$ follows from Eq.~(\ref{eq19}) by replacing $C^\bot_m (t)$
by $\tilde{C}^\bot_m (\tilde{t})$. The scaling relation Eq.~(\ref{eq21}) implies that the
Markov approximation $K_{\perp}^{(s)}(t) \cong \gamma^{(s)}_\bot (L) \delta (t)$ with

\begin{equation} \label{eq22}
\gamma^{(s)}_\bot (L) =\Big[L^4 / t_K(L)\Big] \int\limits^\infty_0 d \tilde{t} \tilde{K} ^{(s)}_\bot (\tilde{t})
\end{equation}
becomes exact for $ L \rightarrow 0$. Consequently, we find an exponentially relaxing solution of  Eq. (\ref{eq17})

\begin{equation} \label{eq23}
C^{(s)}_\bot (t) = C^{(s)}_\bot (0) \exp [-t/\tau^{(s)}_\bot (L)]  \ \ , \ \ C^{(s)}_\bot (0) =(k_BT)^2/2
\end{equation}
with a relaxation time $\tau_{\perp}^{(s)}(L) = 1/\gamma^{s}_\bot (L)$ diverging as $L^{-3}$. Calculating the
integral in Eq.~(\ref{eq22}) one arrives at (see Appendix B)

\begin{eqnarray} \label{eq24}
\tau^{(s)}_\bot (L) \simeq  \frac{1}{8 c}  \Big(\frac{\zeta}{L}\Big)^2
\Big(\frac{L_{a \upsilon}} {L} \Big)^2 t_K (L)  \sim L^{-3}
\end{eqnarray}
with $c \cong 0.019206$ and $L_{a \upsilon}=n_0^{-1/2}$ the average lateral distance between the
particles. The length $\zeta$ characterizes the decay of the pair potential and is defined by
$\zeta^{-2} =(\pi/2) \int\limits^\infty_0 d r r ^{-1} [\upsilon' (r)/(k_BT)]^2 g_{\parallel}(r)$, with $g_{\parallel}(r)$ the pair distribution function of the 2D fluid of the unperturbed lateral d.o.f..

The Mori equation for the correlation function $C_{\parallel}(t)$ describing the relaxation of the energy
of the lateral d.o.f. has the same form as Eq.~(\ref{eq17}) but with a kermel $K_{\parallel}(t)$. As shown
in Appendix C $K_{\parallel} (t)$ is of $O(L^4)$ and decays on a time scale $\tau^{(2D)}$, the
\textit{structural} relaxation time of the unperturbed lateral d.o.f.. The Markov approximation again becomes
exact, for $L \rightarrow 0$. Consequently it follows

\begin{equation} \label{25}
C_{\parallel}(t) = C_{\parallel} (0) \exp [-t/\tau_{\parallel} (L)] , \quad C_{\parallel}(0) =\langle (\delta H_{\parallel})^2 \rangle^{(0)}
\end{equation}
with a relaxation time $\tau_{\parallel}(L)$ diverging as $L^{-4}$. In contrast to $K_\bot ^{(s)}(t)$ the
corresponding integral $\int\limits_0^\infty dt K_{\parallel}(t)$ can not be computed analytically. Hence the prefactor of the $L^{-4}$ divergence can not be calculated exactly.

\section{Summary and Conclusions}\label{Sec: summary-conclusion}

We have shown that strongly confined fluids with Newtonian dynamics  in a slit geometry  possess  interesting features. This is just the situation where a fluid becomes quasi-two-dimensional. The confined (i.e., transversal) and unconfined (i.e., lateral) degrees of freedom (d.o.f.) decouple for vanishing slit-width $L$ and become a Knudsen gas and a 2D fluid, respectively. The structural relaxation of the former exhibit a Gaussian long-time decay with a relaxation time $t_K(L) = 2L/v_{th}$.  $t_K(L)$ is the time for a bounce of a particle with thermal velocity  $v_{th}$. 

For small  but finite  $L$, the coupling between these d.o.f. leads to an exponential decay of  energy fluctuations of the confined and unconfined d.o.f. with diverging relaxation time scales $\tau^{(s)}_{\perp}(L) \sim L^{-3}$ and $\tau_{\parallel}(L) \sim L^{-4}$, respectively. Due to the different power law divergences, the unconfined d.o.f. will equilibrate much slower than the confined ones, provided $L$ is small enough. These results  are only valid for \textit{smooth} pair potentials $v(r)$. If the pair potential becomes more and more hard-core-like, e.g., the relaxation kernel $K^{(s)}_{\perp}(t)$ will gain an increasing contribution at $t=0$ because it involves the derivative of  $v(r)$ (see Appendix B for details). This 
leads to an additional contribution to the damping constant $\gamma^{(s)}_{\perp}(L)$ which may modify its $L$ dependence.  The same may happen for $\gamma_{\parallel}(L)$. In this context
it is also interesting that studying the mode coupling equations \cite{G09} for a confined fluid \cite{LBOHFS10,LSKF12} a divergent time scale has been predicted indirectly from the noncommutativity of the limits $L \to 0$ and $t \to \infty$ \cite{LSF14}.

Usually, diverging time scales are believed to result from a diverging length scale (see, e.g., Ref. \cite{MS06}).  For instance, this is true for the critical fluctuations close to a second order phase transition.  The situation for strongly confined fluids is just opposite. A vanishing confinement length implies  diverging time scales. Or stated otherwise, the planar limit of a fluid is connected with an unlimited increase of the equilibration time. This fact is related to the weak coupling limit between two subsystems, which has been studied over decades \cite{Sp80}. But there are  differences to the earlier investigations. For the strongly confined fluid there are two  coupling constants
$\lambda_{\perp} \sim L$ and $\lambda_{\parallel} \sim L^2$
related to the Liouvillean $\mathcal{L}_1^{\perp}$ and $\mathcal{L}_1^{\parallel}$, respectively. Applying the result from Ref. \cite{vH55} one would predict time scales diverging like $\sim L^{-2}$ and $\sim L^{-4}$. The discrepancy between the former and the exact result (\ref{eq24}) demonstrates that it is not only the coupling constant itself which determines the relaxation time scale   but also the relaxational dynamics within the subsystems.
It would be interesting to search for similar physical situations for which the tuning of a control parameter  leads to weakly coupled subsystems and consequently to diverging time scales.

Our predictions can be  checked  both, experimentally and by MD simulations. At least two possible setups could be used. The first one is a direct approach.  Because $C^{(s)}_{\perp}(t)$ and  $C_{\parallel}(t)$ is the autocorrelation function of the energy  fluctuation of the confined and unconfined d.o.f., respectively,  one could use differential calorimetry. This was worked out analytically \cite{GL89} and applied in order to determine a frequency dependent specific heat by a MD simulation \cite{SKLHB01}.  Second, as already mentioned in the introduction generating a nonequilibrium state by applying an external perturbation, the unconfined and the confined d.o.f. will converge to a local quasi-equilibrium which in turn will relax to the global equilibrium state on much larger time scales $\tau_{\parallel}(L)$ and  $\tau^{(s)}_{\perp}(L)$.
Measuring for instance for different $L$ the 
quasi-static structure factor $S_{\parallel}(q_0,0;t_w)$ of the unconfined d.o.f. at the first peak position $q_0$ as function of the waiting time $t_w$  would allow to determine indirectly $\tau_{\parallel}$ as a function of $L$. Similarly, measuring one of the transversal correlators, e.g., $S_{11}(q,t) \simeq S^{(K)}_{11}(t)$ allows to determine the relaxation time $t_K(L;t_w)$. From its 
dependence on $t_w$ one could deduce the relaxation time $\tau_{\perp}^{(s)}$
as a function of $L$. Such studies would also allow to check the range of validity of the power law divergences of $\tau^{(s)}_{\perp}(L)$ and of  $\tau_{\parallel}(L)$.  In contrast to experiments the realization of flat hard walls for a MD simulation should be straightforward. For an experimental approach it would be necessary to choose large spherical particles such that the roughness of the walls on an atomic length scale does not influence the dynamics.
Let us estimate, e.g.,  $\tau^{(s)}_{\perp}(L)$ for a fluid of argon atoms ($^{40}_{80} Ar$) with Lennard-Jones potential at room temperature. With $\epsilon_{LJ}/k_B \cong 125.7 K$ \cite{W99}, $L=0.1 \zeta_{LJ}$ and choosing the 2D number density such that $L/L_{av}=0.1$ we get from Eq. (\ref{eq24}) that $\tau^{(s)}_{\perp}(0.1 \zeta_{LJ}) \cong 3.53\times 10^5 ~ t_K(0.1 \zeta_{LJ})$. Therefore, the equilibration of the confined (transversal) d.o.f.  requires the particles to bounce back and forth about  a million times. If $L$ would be decreased by a factor of ten the equilibration would  already need about ten billions of bounces.

Finally, we point out that these findings also hold qualitatively for a fluid in a narrow cylindrical tube with radius $R$ and in a narrow two-dimensional channel with width $W$,  replacing $L$ by the radius $R$ and the width $W$, respectively. 
Therefore, it would be interesting performing similar investigations for fluids in narrow tubes and in narrow two-dimensional channels.

\begin{acknowledgments}
Helpful discussions with Thomas Franosch, Simon Lang, Suvendu Mandal and Peter Talkner are gratefully acknowledged. I also would like to thank Thomas Franosch, Tadeus Ras and Peter Talkner for pointing out Refs. \cite{FS86}, \cite{H16} and \cite{Sp80}, respectively.
\end{acknowledgments}

\appendix

\section{Calculation of correlation functions of the Knudsen gas}
\label{sec:Knudsengas}

The one-dimensional Knudsen gas consists of $N$ noninteracting point particles
with mass $m$ confined between two neutral "walls" with positions at $z_{\pm}= \pm L/2$. Its equilibrium phase is described by the canonical ensemble
\begin{align}
\label{eq:Kensemble}
  \rho_K(\{z_i\},\{\dot{z}_i\}) = & \prod_{n=1}^N \frac{\exp(-m \dot{z}_n^2/2 k_B T ) }{L \sqrt{2\pi k_B T /m}}  \nonumber \\
& =
\prod_{n=1}^N \rho_{0}(z_n,\dot{z}_n) \ .
\end{align}

Let $A_{\mu}(z_s)$ be functions depending on a single coordinate $z_s$ and $z_s(t;z_s,\dot{z}_s)$ the Newtonian trajectory of particle $s$ with inital conditions
$(z_s,\dot{z}_s)$. Then the time dependent correlation functions of the observables  $A_{\mu}(z_s)$ is given by  averaging over the transversal degrees of freedom
\begin{eqnarray}
\label{eq:Acorrelator-1}
&&\langle A_{\mu}(t)^{*}A_{\nu}(0) \rangle^{\perp} =\int_{-L/2}^{L/2} \diff z_s \int_{- \infty}^{+\infty}\diff \dot{z}_s \times \nonumber \\
&&
~ \times A_{\mu}(z_s(t;z_s,\dot{z}_s))^{*}A_{\nu}(z_s)~\rho_{0}(z_s,\dot{z}_s) 
\end{eqnarray}
where $ A_{\mu}^{*}$ is the complex conjugate of $A_{\mu}$. This can be rewritten by use of the one-particle distribution function $f(z,\dot{z},t|z_0,\dot{z}_0)$ with initial condition $f(z,\dot{z},0|z_0,\dot{z}_0)= \delta(z-z_0) \delta(\dot{z}-\dot{z}_0)$

\begin{eqnarray}
\label{eq:Acorrelator-2}
&&\langle A_{\mu}(t)^{*}A_{\nu}(0) \rangle^{\perp}= \int_{-L/2}^{L/2} \diff z \int_{-L/2}^{L/2} \diff z_0 \int_{-\infty}^{\infty} \diff \dot{z}\int_{-\infty}^{\infty} \diff \dot{z}_0  \times \nonumber \\
&& \quad \times ~ A_{\mu}(z)^{*}A_{\nu}(z_0)f(z,\dot{z},t|z_0,\dot{z}_0)\rho_{0}(z_0,\dot{z}_0) \ .
\end{eqnarray}
$f(z,\dot{z},t|z_0,\dot{z}_0)$ was calculated analytically \cite{L74}. Taking into account that compared to our geometry the "walls" in Ref. \cite{L74} where shifted by $L/2$  it is
\begin{eqnarray}
\label{eq:Kprobability}
&& f(z-L/2,\dot{z},t| z_0-L/2, \dot{z}_0) = \frac{1}{L} \Big\{ 1+  \nonumber \\ 
&& + 2\sum_{n=1}^\infty \cos( \frac{1}{2} Q_n z_0) \cos[ \frac{1}{2} Q_n (z- \dot{z}_0 t) ] \Big\} \delta(\dot{z}-\dot{z}_0) 
\end{eqnarray}
with the discrete wave numbers $Q_n=2\pi n/L$.

In a first step we will use Eqs. (\ref{eq:Kensemble}),(\ref{eq:Acorrelator-2}) and (\ref{eq:Kprobability}) to calculate the intermediate scattering functions of the Knudsen gas defined by $S_{\mu\nu}^{(K)}(t) = \langle \text{e}^{-\text{i} [ Q_\mu z_s(t)- Q_\nu z_s(0) ] } \rangle^{\perp}$. After shifting the integration varibles $z$ and $z_0$ one obtains with $A_{\mu}(z)= \exp{(iQ_{\mu}z)}$ from Eq. (\ref{eq:Acorrelator-2})
\begin{eqnarray}
\label{eq:SKnudsenintegral}
S_{\mu\nu}^{(K)}(t) &=& (-1)^{\mu+\nu}\int_0^L \!\! \diff z \int_0^L \!\!\diff z_0 \int_{-\infty}^\infty \!\!\diff \dot{z} \int_{-\infty}^\infty \!\!\diff \dot{z}_0  \times \nonumber \\
&& \times \text{e}^{-\text{i} [Q_\mu z-Q_\nu z_0]} f(z-L/2,\dot{z},t|z_0-L/2,\dot{z}_0) \times \nonumber \\
&& \times \rho_0(z_0-L/2,\dot{z}_0)  
\end{eqnarray}
where the factor $ (-1)^{\mu+\nu}$ results from the shift of coordinates.
Substitution of $f(z-L/2,\dot{z},t| z_0-L/2, \dot{z}_0)$ from Eq. (\ref{eq:Kprobability}) and of $\rho_0(z_0,\dot{z}_0)$ from Eq. (\ref{eq:Kensemble}) involves two types of integrals
\begin{align}
\label{eq:integral-1}
\int_{-\infty}^{\infty} \diff \dot{z}_0 \cos(Q_n\dot{z}_0t/2) \exp(-m\dot{z}_0^2/2k_BT) = \nonumber \\
=\sqrt{2 \pi k_B T /m} \exp[-2\pi^2n^2(t/t_K(L))^2]
\end{align}
with $t_K(L)$ from Eq. (\ref{eq15}). The corresponding integral with $\cos(Q_n\dot{z}_0t/2)$ replaced by $\sin(Q_n\dot{z}_0t/2)$  vanishes, because the integrand is an odd function of $\dot{z}_0$. The 2nd integral is as follows
\begin{align}
\label{eq:integral-2}
\int_0^L \diff z \text{e}^{\text{i} Q_\nu z} \cos(\frac{1}{2} Q_n z) = L
\begin{cases}
\frac{1}{2}( \delta_{n,-2\nu} + \delta_{n,2\nu}) \ , & n \text{ even} \\
\frac{2\text{i}}{\pi} \frac{2\nu}{(2\nu)^2 - n^2} \ \ , & n \text{ odd} \ \ .
\end{cases}
\end{align}
Taking these results into account one finally obtains

\begin{eqnarray}
\label{eq:SKnudsen-1}
 S_{\mu\nu}^{(K)}(t) &=& (-1)^{\mu +\nu} \Big\{ \delta_{\mu 0}\delta_{\mu\nu} + \nonumber \\
&&+ \text{e}^{-8 \pi^2 \mu^2 (t/t_K(L))^2 } (1-\delta_{\mu 0}) (\delta_{\mu,-\nu} + \delta_{\mu\nu}) \nonumber \\
&& +
 \sum_{k=0}^\infty c^{k}_{\mu\nu} \text{e}^{- 2 \pi^2 (2k+1)^2 (t/t_{K}(L))^2}\Big\}\ \ .
\end{eqnarray}
with the coefficients

\begin{align}
\label{eq:SKnudsen-2}
 c^{k}_{\mu\nu} =
\frac{8}{\pi^2}  \frac{(2\mu) (2\nu) }{[(2\mu)^2 - (2k +1)^2] [(2\nu)^2 - (2k +1)^2] }   \ .
\end{align}

Next we calculate the correlators $C_m^{\perp}(t)= \langle (z_s(t))^m~(z_s)^m \rangle^{\perp}$ introduced in Sec. IV.  With $A_{\mu}(z_s)=(z_s)^m$ we obtain from Eq. (\ref{eq:Acorrelator-2}) after shift of the integration variables

\begin{eqnarray}
\label{eq:zKnudsenintegral-m}
C_m^{\perp}(t) &=& \int_0^L \!\! \diff z \int_0^L \!\!\diff z_0 \int_{-\infty}^\infty \!\!\diff \dot{z} \int_{-\infty}^\infty \!\!\diff \dot{z}_0  \times \nonumber \\
&& \times (z-L/2)^m (z_0-L/2)^m \rho_0(z_0-L/2,\dot{z}_0) \times \nonumber \\
&& \times f(z-L/2,\dot{z},t|z_0-L/2,\dot{z}_0)  .
\end{eqnarray}
Besides the integral (\ref{eq:integral-1}) it involves elementary integrals
$\int_0^L \diff z z^k \cos(Q_n z/2)$. Using the expressions for those integrals
one gets finally

\begin{align}
\label{eq:zKnudsenintegral-1}
 C_1^{\perp}(t) = & L^2 \frac{8}{\pi^4} \sum_{k=0}^\infty \frac{1}{(2k+1)^4} \text{e}^{- 2 \pi^2 (2k+1)^2 (t/t_{K}(L))^2} \nonumber \\
&
= L^2~ \tilde{C}_1^{\perp}(\tilde{t}) \  .
\end{align}
and

\begin{align}
\label{eq:zKnudsenintegral-2}
 C_2^{\perp}(t) =  & L^4 [\frac{1}{144}+ \frac{8}{\pi^4} \sum_{k=1}^\infty \frac{1}{(2k)^4} \text{e}^{- 2 \pi^2 (2k)^2 (t/t_{K}(L))^2}] \nonumber \\
&
= L^4~ \tilde{C}_2^{\perp}(\tilde{t}) \ .
\end{align}
where $\tilde{t}=t/t_K(L)$  is a dimensionless time. One can prove that the results
(\ref{eq:zKnudsenintegral-1}) and (\ref{eq:zKnudsenintegral-2}) fulfil the correct initial conditions $C_m^{\perp}(0)= \langle z_s^{2m}\rangle^{\perp}=\frac{1}{2m+1}(L/2)^{2m}$.

\section{Calculation of $K_{\perp}^{(s)}(t)$ and of $\gamma_{\perp}^{(s)}$ for $L \to 0$}
\label{sec:kernel-perp}
In the following it is $h_{\perp}(P_s^z)=(P_s^z)^2/2m$ and  $\delta h_{\perp}(P_s^z)=h_{\perp}(P_s^z) - \langle h_{\perp}(P_s^z)\rangle$ its fluctuation. The perturbational approach for $L \to 0$ to calculate the memory kernel $K_{\perp}^{(s)}(t)$  (cf. Eq. (\ref{eq18})) is straightforward and can be done systematically (see, e.g., Ref. \cite{FS86}). Taking into account that $\mathcal{Q}_{\perp}\mathcal{L}\delta h_{\perp}(P_s^z)=\mathcal{L}\delta h_{\perp}(P_s^z) =\mathcal{L}_0\delta h_{\perp}(P_s^z)+\mathcal{L}_1\delta h_{\perp}(P_s^z)=\mathcal{L}_1\delta h_{\perp}$ and $\mathcal{L}_1\delta h_{\perp}=\mathcal{L}^{\parallel}_1\delta h_{\perp}+\mathcal{L}^{\perp}_1\delta h_{\perp}=\mathcal{L}^{\perp}_1\delta h_{\perp}$ where Eqs. (\ref{eq6}) - (\ref{eq9+10}) were applied one obtains from Eq. (\ref{eq18})  in leading order in $L$

\begin{equation}
\label{eq:Kperp-1}
 K^{(s)}_{\perp}(t) \simeq \langle \delta h_{\perp} \mathcal{L}^{\perp}_1 \text{e}^{-i \mathcal{Q}_{\perp}^{(0)}\mathcal{L}_0\mathcal{Q}_{\perp}^{(0)}t}  \mathcal{L}^{\perp}_1  \delta h_{\perp}\rangle^{(0)}/\langle (\delta h_{\perp})^2\rangle^{(0)}.
\end{equation}
where $\mathcal{Q}_{\perp}^{(0)}= 1 - \mathcal{P}_{\perp}^{(0)}$ with

\begin{equation}
\label{eq:projperp-0}
 \mathcal{P}_{\perp}^{(0)}= \Big(|\delta h_{\perp} \rangle^{(0)(0)} \langle \delta h_{\perp}| \Big)/\langle (\delta h_{\perp})^2\rangle^{(0)}
\end{equation}
and $\langle(\cdots)\rangle^{(0)}$ denotes the canonical average of $(\cdots)$
with respect to the unperturbed Hamiltonian $H_0$ obtained from $H$ (Eqs. (\ref{eq1}), (\ref{eq3}) and (\ref{eq4})),
by neglection of the  coupling term $V_{\parallel,\perp}$. Note that $\langle (\delta h_{\perp})^2\rangle^{(0)}=\langle (\delta h_{\perp})^2\rangle=(k_BT)^2/2$. The expression (\ref{eq:Kperp-1}) for the kernel simplifies even more taking into account that $\mathcal{L}_0=\mathcal{L}^{\parallel}_0+\mathcal{L}^{\perp}_0$ and

\begin{equation}
\label{eq:commutator}
[\mathcal{Q}_{\perp}^{(0)},\mathcal{L}^{\parallel}_0] =0 \ \ \ , [\mathcal{Q}_{\perp}^{(0)},\mathcal{L}^{\perp}_0]=0  \ .
\end{equation}
This can be proved by operating with these commutators on phase space functions $f_{\parallel}(\{\vec{r}_i\},\{\vec{P}_i\})g_{\perp}(\{z_i\},\{P^z_i\})$ and using
$\mathcal{L}^{\parallel}_0\delta h_{\perp}(P_s^z) \equiv0$ and $\mathcal{L}^{\perp}_0\delta h_{\perp}(P_s^z) \equiv 0$. Eq. (\ref{eq:commutator}) together with
$(\mathcal{Q}_{\perp}^{(0)})^2 = \mathcal{Q}_{\perp}^{(0)}$ and
$[\mathcal{L}^{\parallel}_0,\mathcal{L}^{\perp}_0] =0$ leads to

\begin{equation}
\label{eq:exponential}
\text{e}^{-i \mathcal{Q}_{\perp}^{(0)}\mathcal{L}_0 \mathcal{Q}_{\perp}^{(0)} t }
= \text{e}^{-i \mathcal{L}^{\parallel}_0  t} \text{e}^{-i \mathcal{L}^{\perp}_0  t}\mathcal{Q}_{\perp}^{(0)} \ .
\end{equation}

Because $\mathcal{Q}_{\perp}^{(0)} \mathcal{L}^{\perp}_1  \delta h_{\perp}(P_s^z) =\mathcal{L}^{\perp}_1  \delta h_{\perp}(P_s^z) - \mathcal{P}_{\perp}^{(0)} \mathcal{L}^{\perp}_1  \delta h_{\perp}(P_s^z)$ and $ \mathcal{P}_{\perp}^{(0)} \mathcal{L}^{\perp}_1  \delta h_{\perp}(P_s^z) \sim \langle \delta h_{\perp} \mathcal{L}^{\perp}_1\delta h_{\perp} \rangle^{(0)} = 0$
(since $\mathcal{L}^{\perp}_1\delta h_{\perp}(P_s^z)$ is an odd function of $P_s^z$ whereas $\delta h_{\perp}(P_s^z)$ is even) we arrive at

\begin{equation}
\label{eq:Kperp-2}
 K^{(s)}_{\perp}(t) \simeq \langle \delta h_{\perp} \mathcal{L}^{\perp}_1 \text{e}^{-i \mathcal{L}^{\parallel}_0t} \text{e}^{-i  \mathcal{L}^{\perp}_0 t} \delta h_{\perp}\rangle^{(0)} 2/(k_B T)^2  \ .
\end{equation}
With $\mathcal{L}^{\perp}_1$ from Eq. (\ref{eq9+10}) we get in leading order in $L$

\begin{equation}
\label{eq:liouville-1-perp}
 \mathcal{L}^{\perp}_1 \delta h_{\perp}(P_s^z) \simeq -(2i/m) \sum_{n (\neq s)} v_1(r_{ns})(z_n-z_s) P_s^z \  .
\end{equation}
Substitution of this expression into Eq. (\ref{eq:Kperp-2}) yields

\begin{eqnarray}
\label{eq:Kperp-3}
 K^{(s)}_{\perp}(t) &\simeq &(4/m^2) \sum_{\small{
\begin{array}{c}m,n\\
(\neq s)
\end{array}}
} \langle v_1(r_{ms}) \text{e}^{-i \mathcal{L}^{\parallel}_0t}  v_1(r_{ns}) \rangle^{\parallel} \times \nonumber   \\
&& \times \langle (z_m-z_s)P_s^z \text{e}^{-i  \mathcal{L}^{\perp}_0 t} (z_n-z_s)P_s^z \rangle^{\perp} \times \nonumber   \\
&& \times2/(k_B T)^2 \ .
\end{eqnarray}
where we used that $\langle \ \rangle^{(0)}= \langle \ \rangle^{\parallel} \langle \ \rangle^{\perp}$. $ \langle \ \rangle^{\parallel}$ is the average with respect to the unperturbed Hamiltonian $H_0^{\parallel}= \sum_{i} \vec{P_i}^2/2m + V(\{\vec{r}_i\})$ and $\langle \ \rangle^{\perp}$ is the average with respect to the unperturbed Hamiltonian $H_0^{\perp}= \sum_{i} [(P_i^z)^2 + \mathcal{U}(z_i)]$
with $\mathcal{U}(z)= \infty$ for $|z| > L/2$ and  zero otherwise. Since $ \mathcal{L}^{\perp}_0$ describes the dynamics of a one-dimensional ideal gas confined to $-L/2 \leq z_i \leq L/2$ (Knudsen gas) the transversal correlator in Eq. (\ref{eq:Kperp-3}) can be expressed  by transversal correlators of the $s$th particle.
Since the particles are identical, and $m \neq s$ and $n \neq s$ one obtains

\begin{eqnarray}
\label{eq: zcorrelatordecomp-1}
&&\langle (z_m-z_s)P_s^z \text{e}^{-i  \mathcal{L}^{\perp}_0 t} (z_n-z_s)P_s^z \rangle^{\perp} = \nonumber \\
&& = \begin{cases}
\langle z_s(t)P_s^z(t)  z_s P_s^z \rangle^{\perp} \ , & m \neq n  \\
\langle z_s(t) z_s \rangle^{\perp} \langle P_s^z(t) P_s^z \rangle^{\perp}+\langle z_s(t)P_s^z(t)  z_s P_s^z \rangle^{\perp} \ \ , & m=n  \ \ .
\end{cases}
\end{eqnarray}
where we used that $\langle z \rangle^{\perp}=0$, due to the symmetric (neutral) wall potential~ $\mathcal{U}(z)$. Introducing the correlators $C_m^{\perp}(t)=\langle (z_s(t))^m (z_s)^m \rangle^{\perp}$ (cf. Appendix A) and taking into account the time reversal symmetry we obtain from Eq. (\ref{eq: zcorrelatordecomp-1})

\begin{eqnarray}
\label{eq: zcorrelatordecomp-2}
&&\langle (z_m-z_s)P_s^z \text{e}^{-i  \mathcal{L}^{\perp}_0 t} (z_n-z_s)P_s^z \rangle^{\perp} = \nonumber  \\
&& =-(m^2/4)
\begin{cases}
\ddot{C}_2^{\perp}(t) \ , & m \neq n  \\
4 C_1^{\perp}(t) \ddot{C}_1^{\perp}(t) + \ddot{C}_2^{\perp}(t) \ \ , & m=n  \ \ .
\end{cases}
\end{eqnarray}
where the dots denote the derivatives with respect to time.
Since the correlators $C_m^{\perp}(t)$ decay on a time scale $t_K(L) \sim L$ which is much faster than the decay of the lateral correlators in Eq. (\ref{eq:Kperp-3}), the latter can be replaced by their initial values at $t=0$.  Taking this and Eq. (\ref{eq: zcorrelatordecomp-2}) into account we get from Eq. (\ref{eq:Kperp-3}) the final result in leading order in $L$

\begin{equation}
\label{eq:Kperp-final}
 K^{(s)}_{\perp}(t) \simeq - k_1 C_1^{\perp}(t) \ddot{C}_1^{\perp}(t) - k_2  \ddot{C}_2^{\perp}(t) \ ,
\end{equation}
which is identical to Eq. (\ref{eq19}).
The coefficients are given by $k_1= 8B/(k_BT)^2$ and $k_2=2(A+B)/(k_BT)^2$
with

\begin{eqnarray}
\label{eq:Kperp-coefficients-1}
A& =& \sum_{\small{
\begin{array}{c}m,n\\
(\neq s)
\end{array}}
}  \langle v_1(r_{ms}) v_1(r_{ns}) \rangle^{\parallel} \nonumber \\
B &=& \sum_{m (\neq s)}   \langle (v_1(r_{ms}))^2  \rangle^{\parallel} \ .
\end{eqnarray}
 By use of the $M$-particle densities $\rho^{(M)}_{\parallel}(\vec{y}_1,\cdots,\vec{y}_M)= (n_0)^M g_{\parallel}^{(M)}(\vec{y}_1-\vec{y}_M, \cdots,\vec{y}_{M-1}-\vec{y}_M)$ \cite{HMcD06} with $ g_{\parallel}^{(M)}$ the $M$-particle distribution function of the two-dimensional fluid of the unperturbed lateral degrees of freedom and $n_0 = N/A$ its number density one gets

\begin{eqnarray}
\label{eq:Kperp-coefficients-2}
A &=& (n_0)^2 \int \diff^2 r \int \diff^2 r' ~ g_{\parallel}^{(3)}(\vec{r},\vec{r'})v_1(r)v_1(r') \nonumber \\
B &=& n_0 \int \diff^2 r ~ g_{\parallel}^{(2)}(r)(v_1(r))^2 \ .
\end{eqnarray}

The damping constant $\gamma_{\perp}^{(s)}= \int_0^{\infty} \diff t K^{(s)}_{\perp}(t)$  follows by substituting the scaling relations for $C^{\perp}_m(t)$ from Eqs. (\ref{eq:zKnudsenintegral-1}) and (\ref{eq:zKnudsenintegral-2}) into Eq. (\ref{eq:Kperp-final}). 
 Then it is straightforward to prove that $\int_0^{\infty} \diff \tilde{t} ~ \ddot{\tilde{C}}_2^{\perp}(\tilde{t})=0$ and $\int_0^{\infty} \diff \tilde{t} ~ \tilde{C}_1^{\perp}(\tilde{t}) \ddot{\tilde{C}}_1^{\perp}(\tilde{t})=- \int_0^{\infty} \diff \tilde{t} ~(\dot{\tilde{C}}_1^{\perp}(\tilde{t}))^2 =-c $ where

\begin{eqnarray}
\label{eq:coefficient-c}
c &=&  (64 \sqrt{2}/\pi^{13/2}) \sum_{k,k'=0}^{\infty}\big \{(2k+1)^2(2k'+1)^2 \times \nonumber \\
&& \times [(2k+1)^2+(2k'+1)^2]^{3/2} \big \}^{-1}
\nonumber \\
\\
&& \cong 0.019206  \ .
\end{eqnarray}

This yields in leading order in $L$ $\gamma_{\perp}^{(s)}(L) \simeq 8cBL^4/((k_BT)^2t_K(L))$
which leads to a relaxation time
\begin{equation}
\label{eq:relax-time}
\tau_{\perp}^{(s)}(L) \simeq \frac{1}{8c} \big (\frac{\zeta}{L} \big )^2 \big (\frac{L_{av}}{L} \big )^2 ~ t_K(L) \ ,
\end{equation} with $\zeta$ a length characterizing the decay of the pair potential $v(r)$. It is defined by

\begin{equation}
\label{eq:length-zeta}
\zeta^{-2} = \frac{\pi}{2} \int_0^{\infty} \diff r ~  r^{-1} \big [v'(r)/(k_BT) \big]^2 g_{\parallel}^{(2)}(r)    \ .
\end{equation}
 $L_{av}= n_0^{-1/2}$ is the average lateral distance of the particles. Eq. (\ref{eq:relax-time}) is identical to Eq. (\ref{eq24}). Note, in Sec. IV we 
skipped the superscript at the pair distribution function $ g_{\parallel}^{(2)}(r)$.

\section{Calculation of $K_{\parallel}(t)$ and of $\gamma_{\parallel}$ for $L \to 0$}
\label{sec:kernel-parallel}

Similar to Eq. (\ref{eq18}) the kernel $K_{\parallel}(t)$ describing the relaxation of the lateral correlator $C_{\parallel}(t)$ is given by

\begin{equation}
\label{eq:Kpara-0}
 K_{\parallel}(t) = \langle \delta H_{\parallel} \mathcal{L}\mathcal{Q}_{\parallel} \text{e}^{-i \mathcal{Q}_{\parallel}\mathcal{L} \mathcal{Q}_{\parallel}t} \mathcal{Q}_{\parallel} \mathcal{L}  \delta H_{\parallel}\rangle /\langle (\delta H_{\parallel})^2\rangle.
\end{equation}
where $\mathcal{Q}_{\parallel} = 1 - \mathcal{P}_{\parallel}$ and
$\mathcal{P}_{\parallel}$ is the projector onto $\delta H_{\parallel}$, the fluctuation of the total energy of the unperturbed lateral degrees of freedom.
Following the same steps as in Appendix B we obtain in leading order in $L$
\begin{equation}
\label{eq:Kpara-1}
 K_{\parallel}(t) \simeq \langle \delta H_{\parallel} \mathcal{L}_1^{\parallel}
\text{e}^{-i \mathcal{L}^{\parallel}_0t} \text{e}^{-i  \mathcal{L}^{\perp}_0 t}
\mathcal{L}_1^{\parallel} \delta H_{\parallel}\rangle^{(0)} /\langle (\delta H_{\parallel})^2\rangle^{(0)} \ .
\end{equation}
Using $\mathcal{L}_1^{\parallel}$ from Eq. (\ref{eq9+10}) we get in leading order in $L$

\begin{equation}
\label{eq:liouville-1-para}
 \mathcal{L}^{\parallel}_1 \delta H_{\parallel} \simeq (i/m) \sum_{k \neq l} \big [v_1'(r_{kl})/r_{kl}\big ] \vec{r}_{kl} \cdot \vec{P}_{k}(z_k-z_l)^2  \ .
\end{equation}
Substitution of this expression into Eq. (\ref{eq:Kpara-1}) and observing that the t-dependent correlator in that equation factorizes we arrive at

\begin{eqnarray}
\label{eq:Kpara-2}
&& K_{\parallel}(t) \simeq   (1/m^2) \times  \nonumber \\
&& \sum_{k \neq l} \sum_{m \neq n} \langle  \big [v_1'(r_{kl})/r_{kl}\big ] \vec{r}_{kl}  \cdot \vec{P}_{k} ~ \text{e}^{-i \mathcal{L}^{\parallel}_0t} ~ \big [v_1'(r_{mn})/r_{mn} \big ] \vec{r}_{mn}  \cdot \vec{P}_{m} \rangle^{\parallel} \times \nonumber \\
&& \times \langle (z_k-z_l)^2 ~  \text{e}^{-i  \mathcal{L}^{\perp}_0 t} (z_m-z_n)^2 \rangle^{\perp}
/\langle (\delta H_{\parallel})^2 \rangle^{\parallel}  \ .
\end{eqnarray}
which corresponds to Eq. (\ref{eq:Kperp-3}). However, the transversal correlator
$\langle (z_k-z_l)^2 ~  \text{e}^{-i  \mathcal{L}^{\perp}_0 t} (z_m-z_n)^2 \rangle^{\perp}$ differs from the transversal correlator in  Eq. (\ref{eq:Kperp-3}).
The $z$ coordinates appear in a quartic form instead of a quadratic form in  Eq. (\ref{eq:Kperp-3}). Therefore it is of $\mathcal{O}(L^4)$. This transversal correlator again can be expressed by $C_m^{\perp}(t)$, introduced in Appendix A. One obtains

\begin{eqnarray}
\label{eq:Kpara-z-corellator}
&&\langle (z_k-z_l)^2 ~  \text{e}^{-i  \mathcal{L}^{\perp}_0 t} (z_m-z_n)^2 \rangle^{\perp}= \nonumber\\ 
&&=C_2^{\perp}(t) \big[\delta_{km}+\delta_{kn}+\delta_{lm}+\delta_{ln} \big] + \nonumber\\
&&+ (\langle z^2\rangle^{\perp})^2 \big[(1-\delta_{km})+(1-\delta_{kn})+(1-\delta_{lm})+(1-\delta_{ln}) \big]+ \nonumber\\
&&+
4(C_1^{\perp}(t))^2 \big[\delta_{km}\delta_{ln}+\delta_{kn}\delta_{lm} \big]
\end{eqnarray}
which decays  to a nonzero value $(L^4/144)\big[\delta_{km}+\delta_{kn}+\delta_{lm}+\delta_{ln} \big] + (\langle z^2\rangle^{\perp})^2\big[(1-\delta_{km})+(1-\delta_{kn})+(1-\delta_{lm})+(1-\delta_{ln}) \big] \equiv L^4/36$ for $t \to \infty$. Here we used $\langle z^2\rangle^{\perp}= L^2/12$ and Eq.~(\ref{eq:zKnudsenintegral-2}). Consequently, in leading order in $L$ it is the decay of the lateral correlator in Eq. (\ref{eq:Kpara-2}) which is responsible for the damping mechanism. This decay happens on the structural relaxation time scale $\tau^{(2D)}$ of the unperturbed
lateral fluid. Since the amplitude of the kernel $K_{\parallel}(t)$ is of $\mathcal{O}(L^4)$ we have
\begin{equation}
\label{eq:Kpara-scaling}
K_{\parallel}(t)=L^4 \tilde{K}_{\parallel}(\tilde{t}')  \ ,  \ \tilde{t}'=t/\tau^{(2D)}
\end{equation}
which makes again the Markov approximation exact for $L \to 0$. Therefore $C_{\parallel}(t)$ decays exponentially with a relaxation time $\tau_{\parallel}(L) =1/\gamma_{\parallel}(L)$
where
\begin{equation}
\label{eq:damping-para}
\gamma_{\parallel}(L) = L^4 \tau^{(2D)} \int_0^{\infty} \diff \tilde{t} \tilde{K}_{\parallel}( \tilde{t}) ~  \sim L^{4}\ .
\end{equation}
The integral in Eq. (\ref{eq:damping-para}) can not be  calculated analytically.

\end{document}